\documentclass{vgtc}                          %

\graphicspath{{figures/}{pictures/}{images/}{./}} %

\usepackage{times}                     %

\usepackage{tabu}                      %
\usepackage{booktabs}                  %
\usepackage{lipsum}                    %
\usepackage{mwe}                       %

\usepackage{mathptmx}                  %

\usepackage{amsmath}
\usepackage{amssymb}

\onlineid{1131}

\vgtccategory{Research}

\vgtcinsertpkg

\title{Chronotome: Real-Time Topic Modeling for Streaming Embedding Spaces}

\author{
\parbox[t]{1.2in}{\centering Matte Lim\thanks{First two authors contributed equally. Emails: \{catherineyeh, wattenberg, fernanda\}@g.harvard.edu, \{mattelim, pmichala\}@gsd.harvard.edu.}\\ {\scriptsize Harvard University}} \hspace{0.05in}
\parbox[t]{1.2in}{\centering Catherine Yeh$^\ast$\\ {\scriptsize Harvard University}} \hspace{0.05in}
\parbox[t]{1.3in}{\centering Martin Wattenberg\\ {\scriptsize Harvard University}\\ {\scriptsize Google Research}} \hspace{0.05in}
\parbox[t]{1.3in}{\centering Fernanda Vi\'egas\\ {\scriptsize Harvard University}\\ {\scriptsize Google Research}} \hspace{0.05in}
\parbox[t]{1.2in}{\centering Panagiotis Michalatos\\ {\scriptsize Harvard University}}
}

\teaser{
  \centering
  \includegraphics[width=\linewidth]{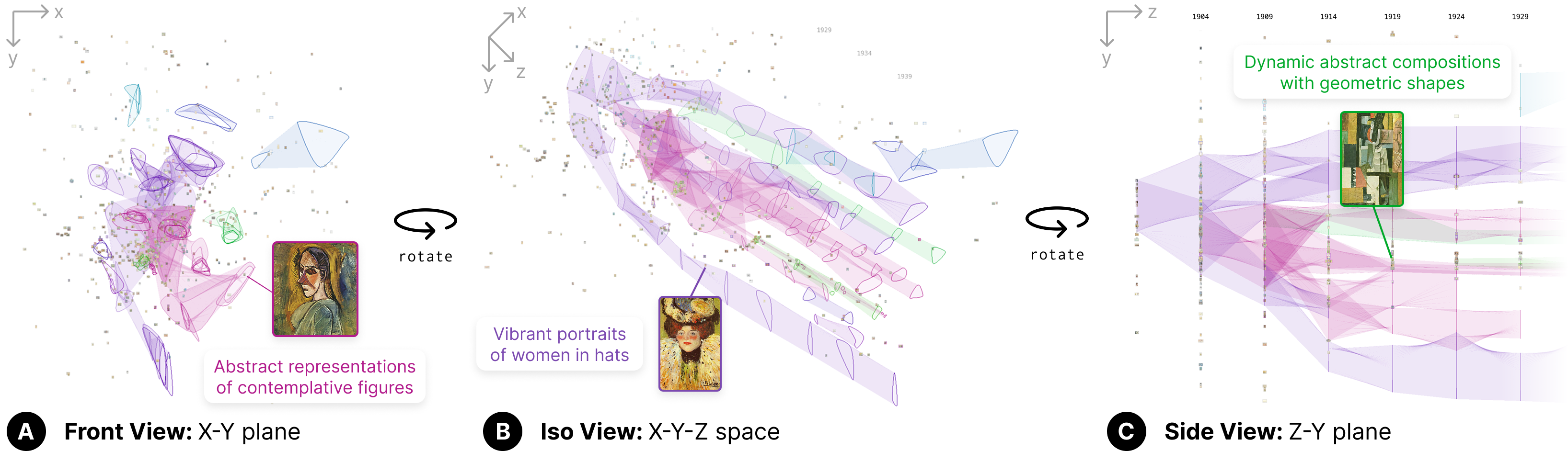}
  \caption{To visualize how topics evolve in real time, we create a rotatable embedding space where time is encoded along the Z-axis. We provide three preset views to help users explore topic clusters from different perspectives: \textbf{(A) Front View} (overall clusters), \textbf{(B) Iso View} (clusters over time), and \textbf{(C) Side View} (clusters over time). Here, each point represents an image from a dataset of Picasso's paintings, batched into 5-year intervals.}
  \label{fig:teaser}
}

\abstract{
    Many real-world datasets -- from an artist's body of work to a person's social media history -- exhibit meaningful semantic changes over time that are difficult to capture with existing dimensionality reduction methods. 
    To address this gap, we introduce a visualization technique that combines force-based projection and streaming clustering methods to build a spatial-temporal map of embeddings. 
    Applying this technique, we create \system, a tool for interactively exploring evolving themes in time-based data -- in real time. 
    We demonstrate the utility of our approach through use cases on text and image data, showing how it offers a new lens for understanding the aesthetics and semantics of temporal datasets.
} %


\begin{document}

\newcommand{\ie}{{i.e.,}\xspace}
\newcommand{\eg}{{e.g.,}\xspace}
\newcommand{\ea}{{et~al\xperiod}\xspace}
\newcommand{\aka}{{a.k.a.}\xspace}
\newcommand{\etc}{{etc\xperiod}\xspace}
\newcommand{\etal}{{et al\xperiod}\xspace}
\newcommand{\vs}{{vs.}\xspace}

\newcommand{\todo}[1]{{\textcolor{OrangeRed}{[Todo: #1]}\normalfont}}
\newcommand{\cy}[1]{{\textcolor{RoyalBlue}{[CY: #1]}\normalfont}}

\newcommand{\system}{\textsc{Chronotome}\xspace} %

\definecolor{DarkPurple}{HTML}{7B49B0}
\definecolor{DarkPink}{HTML}{B21B8C}
\definecolor{DarkGreen}{HTML}{01A92A}
\definecolor{DarkBlue}{HTML}{002AA9}

\definecolor{BillGreen}{HTML}{77BA34}
\definecolor{BillBlue}{HTML}{5747C1}
\definecolor{BillYellow}{HTML}{B69327}
\definecolor{BillRed}{HTML}{B72929}
\definecolor{BillPink}{HTML}{BB379A}

\definecolor{BPPurple}{HTML}{6C46C0}
\definecolor{BPGreen}{HTML}{A1C554}
\definecolor{BPYellow}{HTML}{B88C2F}
\definecolor{BPRed}{HTML}{BF4242}

\newcommand{\coloredsquare}[1]{{\textcolor{#1}{$\blacksquare$}}}

\firstsection{Introduction}

\maketitle

Real-world datasets, such as an artist's evolving body of work~\cite{jones1997influential} or a person's social media timeline~\cite{malik2013topic,schreck2012visual}, often reveal meaningful semantic changes over time. 
Dimensionality reduction (DR) and clustering techniques are a powerful way to visualize contextual, topical relationships, however, they are typically designed for static, one-shot embeddings and cannot easily capture the rich dynamics of temporal data~\cite{bamler2017dynamic,rauber2016visualizing}.
Thus, our work is motivated by the question: How might we create a visual, interpretable map of embeddings that reflects how a dataset evolves over time?

Toward this goal, we design a visualization technique that integrates force-based projection with streaming clustering to build a 3D spatial-temporal embedding map.
Unlike traditional DR approaches, which tend to either support progressive data inclusion (\eg incremental PCA~\cite{artac2002incremental}) or produce more coherent layouts (\eg UMAP~\cite{mcinnes2018umap}, t-SNE~\cite{van2008visualizing}), our method aims to provide both interpretability in structure and continuity over time.
A key aspect of our approach is encoding time along the Z-axis, enabling a rotatable embedding space that visualizes overarching data trends and fine-grained, time-dependent trajectories (Fig.~\ref{fig:teaser}).

We apply this technique to create \system, an interactive tool for real-time exploration of evolving patterns, themes, and structures in temporal datasets. 
Through use cases on text and image data, including Bill Gates' tweets and Picasso's paintings, we showcase how \system can help users surface insights that unfold across time, and point to interesting relationships between local and global semantic trends.

\section{Related Work}
\noindent \textbf{Temporal data visualization.} Researchers use various techniques to visualize data over time~\cite{ali2019clustering}.
For example, dynamic network visualizations are often used to depict relationship-based data such as social networks, transportation routes, and neuron connections~\cite{moody2005dynamic,van2015reducing}. 
Sankey diagrams and parallel coordinate plots are other popular methods for visualizing the flow of multiple categories or variables over time~\cite{malik2013topic,palamarchuk2024visualizing,rodrigues2020cluster}.
More organic illustrations of topic flow can be created using stacked visualizations such as ThemeRiver~\cite{cui2011textflow,havre2002themeriver,liu2012tiara} or Streamgraphs~\cite{byron2008stacked}.

\noindent \textbf{(Incremental) dimensionality reduction.}
Compared to the temporal visualization approaches above, which focus more on capturing trends in quantitative or categorical data~\cite{rodrigues2020cluster,van2015reducing}, dimensionality reduction techniques provide a powerful way of distilling rich, semantic insights from high-dimensional data to visualize in low-dimensional spaces~\cite{artac2002incremental,espadato2021toward,mcinnes2018umap,van2008visualizing} -- especially when combined with clustering techniques (\eg K-means~\cite{macqueen1967some}, DBSCAN~\cite{ester1996density}, HDBSCAN~\cite{mcinnes2017hdbscan}).
DR also allows discovery of new, latent data structures, rather than relying on pre-defined themes.

However, because conventional DR methods like UMAP~\cite{mcinnes2018umap} and t-SNE~\cite{van2008visualizing} are designed to create static embedding plots, it can be difficult to capture dynamic relationships over time~\cite{bamler2017dynamic,rauber2016visualizing}. 
Some work has proposed more streaming-based approaches to DR, such as incremental PCA~\cite{artac2002incremental}, to allow adding new data progressively, but these methods tend to produce less interpretable and coherent visualizations.
We draw inspiration from Parametric UMAP \cite{sainburg2021parametric} as an approach to DR that satisfies both temporal and interpretability goals. 
Still, it is not a real-time process that can be witnessed and inspected visually by users, which we aim to address.

\noindent \textbf{Dynamic topic modeling.}
Dynamic topic modeling is a family of NLP methods that capture how topics in text data evolve, which are often used to support temporal visualizations~\cite{vayansky2020review}.
For example, Latent Dirichlet Allocation (LDA)~\cite{blei2003latent} is one popular technique that uses a Bayesian approach to discover latent topics~\cite{liu2012tiara,malik2013topic}.
Newer methods such as BERTopic~\cite{grootendorst2022bertopic} combine embeddings from pre-trained transformer models, clustering methods, and class-based TF-IDF~\cite{sparck1972statistical} to generate more coherent topic representations.
Researchers have also created streaming approaches to allow incremental, real-time topic modeling (\eg Online LDA~\cite{hoffman2010online}, BERTrend~\cite{boutaleb2024bertrend}), which we take inspiration from in this work.

\section{Technique}
Our visualization technique involves performing force-based dimensionality reduction in the X-Y plane, projecting time onto the Z-axis, and incrementally adding and clustering batches of data.
We use a force-based approach because it enables (1) real-time visualization and understanding of DR processes, and (2) fast, flexible embedding manipulation and experimentation~\cite{kobourov2012spring}.

As input, we take data tuples of the form $x_i=(t_i,o_i,e_i)$, where $t_i$ is a timestamp, $o_i$ is the data object (\eg text or image), and $e_i$ is its vector embedding. 
The output is a graph $G$, where each node corresponds to a data point $x_i$ and is embedded in 3D space. 
$G$'s layout evolves dynamically to reflect changes in content and time.

\subsection{Dimensionality Reduction in X-Y Plane}

We construct a graph $G$ of the data using a spring force model to resolve the positions of nodes in the X-Y plane, similar to classical force-directed approaches~\cite{kobourov2012spring,van2015reducing}.
Every node is connected by an edge, with spring forces that encourage the spatial distance between nodes to match their semantic distance.

Specifically, for each node pair $(i,j)$, we define a target distance $d_{\text{ideal}}(i,j) = 1 - s(i,j)$, where $s(i,j)$ is the cosine similarity between embeddings $e_i$ and $e_j$, and spring constant $k(i,j) = s(i,j)$, which makes semantically similar nodes exert stronger forces.
For each edge, the force applied follows Hooke's Law~\cite{rychlewski1984hooke}:
$$
F_{ij} = k(i,j) \cdot (d_{\text{ideal}}(i,j) - d_{\text{current}}(i,j))
$$
where $d_{\text{current}}(i,j)$ is the current Euclidean distance between nodes $i$ and $j$.
This idea is equivalent to \textit{metric multidimensional scaling}~\cite{kruskal1964multidimensional}, where gradient descent is performed to minimize the squared error of the real and projected distance between pairs of points.
The system minimizes overall energy by iteratively applying a Newtonian move function that updates node positions in real time based on the sum of forces acting upon them.

To gain a better signal for DR, a threshold $\tau$ is applied when calculating the edge forces on nodes. 
$\tau$ acts as a filter such that only node pairs with a high enough cosine similarity $s(i,j) > \tau$ generate attractive spring forces. 
Otherwise, a high repulsive force is applied. 
$\tau$ is dynamically adjusted with $\tau = \mu + (\log N / \log C)\cdot \sigma$, where $\mu$ and $\sigma$ are the mean and standard deviation of all edge similarities, $N$ is the number of nodes, and $C$ is a tunable constant (currently set to $
ln(21)$ based on experimentation). 
This dynamic filtering relaxes constraints imposed by less relevant edges and prioritizes semantically meaningful ones.
Thus, outlier nodes that fall below the threshold are pushed towards the perimeter of the graph.

\subsection{Time Projection Onto Z-Axis}
To incorporate temporal information, we project time onto the Z-axis.
Data is batched based on a user-specified timestep $T$ (\eg 3 months), with each batch corresponding to a discrete slice of time (\eg January 1, 2025 to March 31, 2025). 
Nodes in the same batch are assigned the same Z-coordinate. 
For instance, a node with timestamp $t_i =$ February 20, 2025 would be assigned to the January – March 2025 batch and placed at the corresponding Z-level.

In this way, our approach enables simultaneous exploration of both semantic and structural relationships (X–Y plane) and temporal data progression (Z-axis).

\subsection{Incremental Data Visualization}
When a data batch is added, new nodes and edges are created in the graph. 
As before, the spring constant $k(i,j)$ for each edge is computed upon node addition and proportional to the cosine similarity between embeddings.
Then, the move function resolves the X-Y location of each node.
Each node $i$ has mass $m_i$, which controls how strongly it responds to force, helping to stabilize the graph layout as points are added.
The mass of prior nodes is also scaled by multiplier $\beta$ as new batches arrive to minimize sudden global position changes, with $m_i = m_0 \cdot \beta^{b_{\text{current}}-b_{\text{initial}}}$, where $m_0$ is the original node mass, $b_{\text{initial}}$ is the batch index when the node was first added, and $b_{\text{current}}$ is the current batch index.
We set $\beta=1.618$ (golden ratio), chosen empirically to balance node stability and reactivity.

With this approach, nodes become more resistant to movement over time, 
but are not fixed as their positions remain responsive to the forces introduced by new data. 
This ensures that the layout evolves smoothly and dynamically,
as each node's position reflects both its semantic similarity to other nodes and the graph's structural arrangement at the timestep it was added.

\subsection{Clustering and Graph Evolution}
At each timestep, we perform clustering on the X-Y 2D embedding plane to reveal latent structures.
Clusterings are stored per timestep in order to visualize how node positions and clusters shift over time.
This enables the various views of the graph in Sec.~\ref{sec:interface}. 
Since the dataset grows monotonically, we also track each cluster's ``parent cluster'' from the previous timestep to support continuity. 
This enables cluster splits to be visualized as forked subclusters in later timesteps.
Implementation details are provided in Sec.~\ref{sec:implementation}.

\subsection{Limitations}\label{sec:limitations}

One limitation of using a fully connected graph is that edges scale at $O(n^2)$ times the number of nodes (see Sec.~\ref{sec:implementation} for stats), which puts a computational ceiling on our current method ($n\approx1000$ points).
Improving scalability and supporting cluster merges (in addition to splits) are both important next steps.

\section{\system Design}
Applying our technique, we created \system, an interactive visualization tool for exploring how datasets evolve over time.

\begin{figure*}
    \centering
    \includegraphics[width=\linewidth]{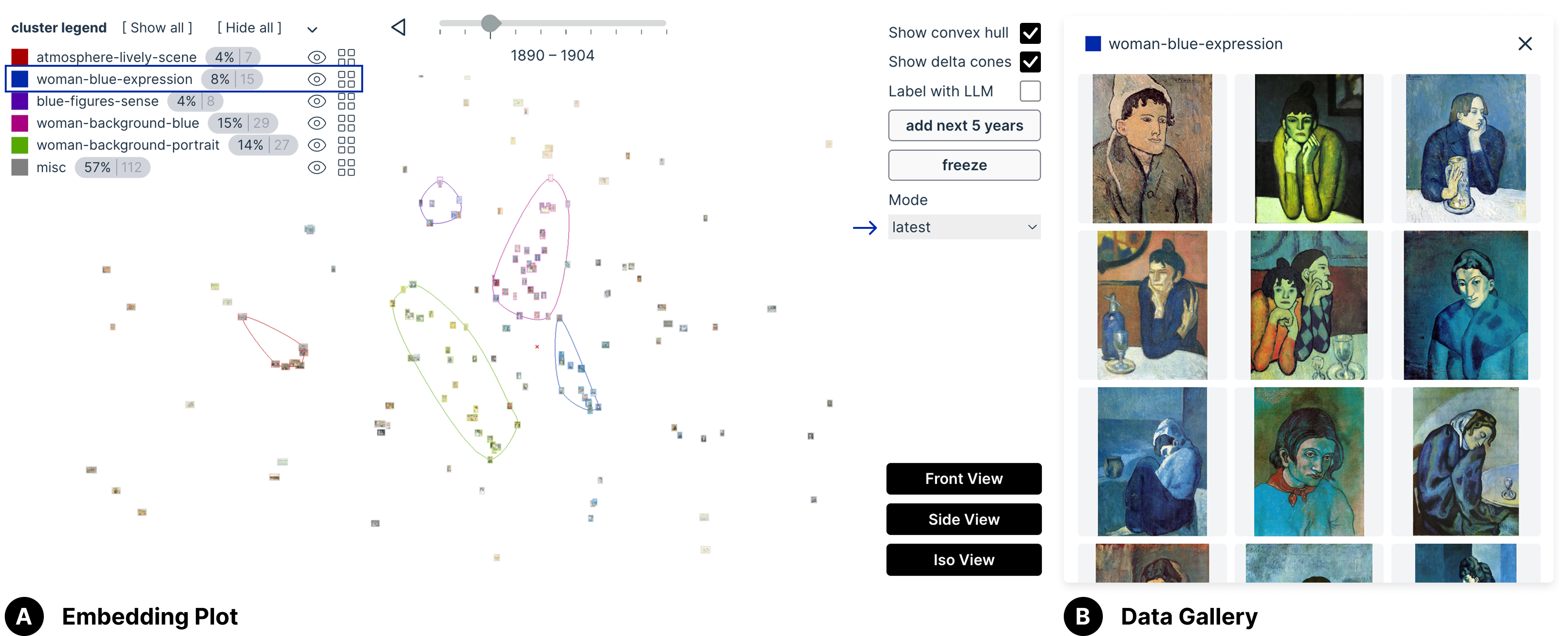}
    \caption{\system contains two main interface components. \textbf{(A)} The rotatable \textbf{Embedding Plot} allows users to explore topic clusters in real time. Currently, the interface is in \textit{Playback} mode, where users can visualize how nodes moved over time via the timestep slider. There is a filterable legend on the left, and additional visualization controls on the right. \textbf{(B)} Clicking on a cluster label in the legend opens the \textbf{Data Gallery}, where users can view all data objects (\eg text or images) in that cluster.}
    \label{fig:interface}
\end{figure*}

\subsection{Interface} \label{sec:interface}

\subsubsection{Embedding Plot}
The Embedding Plot contains our rotatable embedding visualization, where users can explore topic clusters over time (Fig.~\ref{fig:interface}A). 
Each node in the embedding represents an object (\eg text or image) from the current dataset (\eg Picasso's paintings).
Images are visualized directly, while text objects are represented by points.
Node sizes are adjustable to control visual clutter.

On the left, there is a filterable \textit{cluster legend}, which displays topic labels and the proportion of the dataset represented by each cluster.
Points categorized as ``misc'' do not meaningfully belong to any cluster at the current timestep.
Colors are assigned randomly to new clusters, but child clusters inherit color from their parent.

On the right, we provide controls to customize the visualization and add new data points.
Here, the user can toggle the visibility of convex hulls (cluster outlines at each timestep) and delta cones (3D tunnels connecting convex hulls).
If the ``Label with LLM'' checkbox is selected, \system labels clusters using an LLM; otherwise, we use TF-IDF~\cite{sparck1972statistical} (see Sec.~\ref{sec:implementation}).
Clusterings are automatically generated and updated as new data is added.

\noindent \textbf{Interactive modes.}
We include three interactive modes for analyzing the Embedding Plot. 
\textit{Latest} is the default mode where the graph is actively being updated using the move function, which shows the latest node positions and clusterings projected across all timesteps. 

\textit{Across} compares how nodes have shifted over time, as well as how clusters relate in terms of parent-child relationships (Fig.~\ref{fig:teaser}). 
This is the most informationally dense mode as it aims to visualize node movement and cluster evolution within a single view.

\textit{Playback} offers a middle ground between Latest and Across (Fig.~\ref{fig:interface}). 
Users can explore how nodes and clusters changed at each saved timestep via the slider at the top of the interface.

\noindent \textbf{Exploration views.}
At the bottom right corner, we provide three preset views for exploring the embedding space.
\textit{Front View} rotates the graph so the user is looking at the X-Y plane (along the Z-axis) (Fig.~\ref{fig:teaser}A).
This can be useful for visualizing overall trends across the dataset and mirrors a traditional 2D embedding map.

On the other hand, \textit{Iso View} displays depth across all three axes -- X, Y, and Z -- simultaneously (Fig.~\ref{fig:teaser}B), and \textit{Side View} rotates the embedding to look at the Z-Y plane (along the X-axis) (Fig.~\ref{fig:teaser}C).
These views facilitate exploration along the temporal dimension and provide more of a ``parallel coordinates'' inspired view.

\subsubsection{Data Gallery}
Clicking on a topic label in the cluster legend opens the Data Gallery, which shows all the data objects in the corresponding cluster to help users better understand its semantic content (Fig.~\ref{fig:interface}B).

\subsection{Implementation}\label{sec:implementation}
\system is web app with a Typescript/Svelte frontend and a Python/Flask backend.
Node positions are resolved in real time using a Typescript physics engine.
Our effective framerate is 30 fps at 200 nodes, 8 fps at 360 nodes, and 1 fps at 900 nodes. 
Graph force computations and edge creation are offloaded to Web Workers so rendering and interactivity are not impacted by layout updates.
We are working to further parallelize graph resolution with PyTorch.

We create our 3D embedding visualization with \href{https://threejs.org/}{Three.js}, and use \href{https://platform.openai.com/docs/models/gpt-4o-mini}{\texttt{gpt-4o-mini}} as our LLM for cluster labeling.
Currently, we pre-compute the embedding vectors for each dataset, using \href{https://huggingface.co/sentence-transformers/all-MiniLM-L6-v2}{\texttt{all-MiniLM-L6-v2}} -- a sentence transformers model -- for text data, and OpenAI's \href{https://github.com/openai/CLIP}{CLIP} model~\cite{radford2021learning} for image data.
Our system is highly extensible to different modalities (\eg text, image) and supports any form of time-stamped data.

\noindent \textbf{Clustering.}
We use HDBSCAN~\cite{mcinnes2017hdbscan}, a hierarchical method that identifies relatively stable, interpretable clusters of varying densities without requiring the number of clusters to be specified in advance. 
This flexibility is well-suited to dynamic datasets where cluster structure may change over time.
HDBSCAN's robustness to noise also helps isolate outliers, aligning well with our force-directed layout where such nodes drift toward the periphery.

\noindent \noindent \textbf{Topic modeling.}
We implement two topic modeling approaches.
The first is using \textit{TF-IDF}~\cite{sparck1972statistical}, a popular NLP metric used to evaluate word importance within a corpus.
TF-IDF scores help identify the top $m$ most significant terms within each cluster, which are used to represent the topic.
The second method involves sampling text data from each cluster and asking an \textit{large language model (LLM)} summarize the texts with a short, interpretable label.

For text datasets, we pass in sentences directly to generate topic labels.
For image datasets, we first annotate each image with an LLM-generated description of its content.
Then, topic modeling is performed on these natural language descriptions.

\section{Results}
We illustrate \system in action on three datasets.

\noindent \textbf{Bill Gates' tweets (text).} 
We retrieved $n=669$ tweets from Bill Gates' X profile posted from August 31, 2022 to August 21, 2024 using the \href{https://developer.x.com/en/docs/x-api}{Twitter 2.0 API}. 
Only the text and date data was extracted from each tweet to use in our tool. 

\noindent \textbf{Picasso's paintings (image).} 
This dataset ($n=761$) was curated from the \href{https://www.kaggle.com/datasets/steubk/wikiart}{WikiArt dataset} on Kaggle by taking all paintings by Picasso with an annotated year (ranging from 1890 to 1967).
Similarly, we extracted only the date and image data for each painting. 

\noindent \textbf{\textit{The Grand Budapest Hotel} frames (image).}
Our final dataset consists of $n=1000$ frames and time data from the 2014 film, \textit{The Grand Budapest Hotel}, taken from Kaggle's \href{https://www.kaggle.com/datasets/asaniczka/movie-identification-dataset-800-movies}{Movie Identification Dataset}.
The frames are sampled evenly from the movie, which has a run time of 1 hour 39 minutes (\ie in $\sim6.4$ second increments).

\subsection{Bill Gates' Tweets}
\begin{figure}
    \centering
    \includegraphics[width=\linewidth]{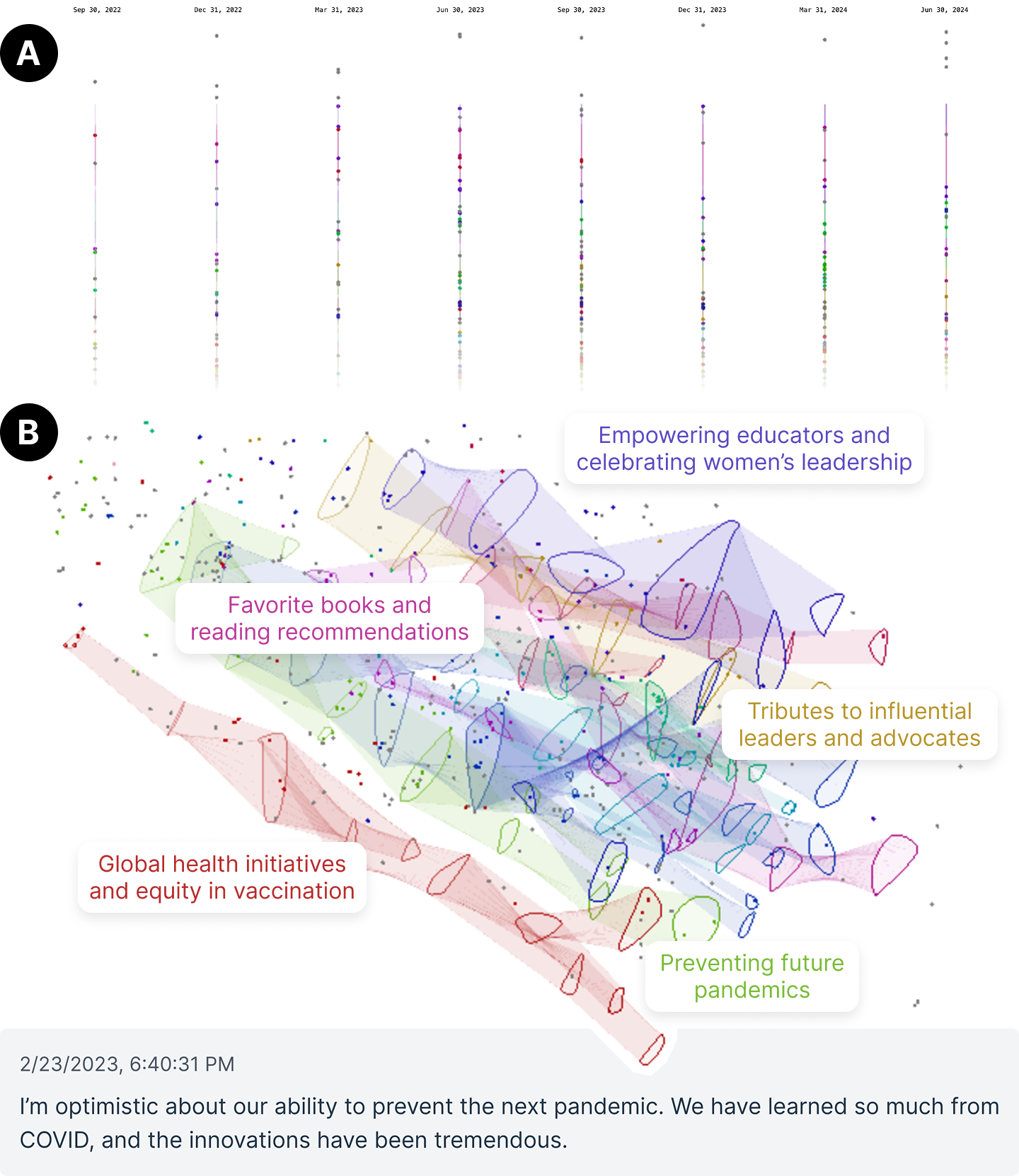}
    \caption{Topic clusters derived from Bill Gates' tweets visualized in 3-month intervals using \textbf{(A)} Side View and \textbf{(B)} Iso View. A example tweet from \coloredsquare{BillGreen} ``Preventing future pandemics'' is shown.}
    \label{fig:bill_gates}
\end{figure}
Carly, a social media analyst, is interested in uncovering temporal trends from Bill Gates' tweets.
She loads the full dataset into \system, using \textit{Across} and \textit{Side View} to explore.
Carly observes that Gates is a fairly consistent tweeter, as each timestep has a similar number of tweets (Fig.~\ref{fig:bill_gates}A).
Switching to \textit{Iso View}, she notices several recurrent themes like \coloredsquare{BillBlue} ``Empowering educators and celebrating women's leadership,'' \coloredsquare{BillRed} ``Global health initiatives and equity in vaccination,'' and \coloredsquare{BillGreen} ``Preventing future pandemics'' (Fig.~\ref{fig:bill_gates}B).
Examining clusters with the Data Gallery, Carly discovers an overall positive tone in Gates' tweets, which are focused on empowerment, optimism about the future, and celebrating others.

Using \textit{Playback}, Carly traces how \coloredsquare{BillGreen} ``Preventing future pandemics'' evolves over time.
The TF-IDF topic labels reveal that this cluster originated from a Covid-specific topic, ``pandemic-covid-world,'' then branched into child clusters like ``omicron-world-pandemic'' and ``covid-19-vaccines,'' before shifting toward broader themes of future pandemic prevention (\eg ``pandemic-prevent-conversation'').
Carly concludes that Gates has a consistent, recognizable online persona, explaining his strong social media impact.
She also appreciates how he frequently incorporates his personal interests such as \coloredsquare{BillPink} ``Favorite books and reading recommendations.''

\subsection{Picasso's Paintings}
Sam is an art student who wants to use \system to learn about their favorite artist, Pablo Picasso. 
Starting in \textit{Latest} and \textit{Front View}, Sam adds a few timesteps of data and finds that Picasso painted fairly realistic works in his early career, including a cluster of horse portraits (``horse-sky-calm'').
They also see Picasso's famous blue period (1901-1904) emerge (\eg~\coloredsquare{DarkBlue} ``woman-blue-expression''), which Sam confirms with the Data Gallery (Fig.~\ref{fig:interface}).

Sam adds more data and tries labeling clusters with an LLM.
Switching to \textit{Across} and \textit{Side View}, Sam observes many persistent themes throughout Picasso's works such as \coloredsquare{DarkPink} ``Emotional and abstract artistic representations,'' which branches into several child clusters, including ``Abstract representations of contemplative figures'' (Fig.~\ref{fig:teaser}).
Similarly, \coloredsquare{DarkPurple} ``Artistic representations of women and femininity'' splits into new clusters like ``Vibrant portraits of women in hats.''
Around 1907-1908, Sam sees a shift toward cubism, which manifests in clusters such as \coloredsquare{DarkGreen} ``Dynamic abstract compositions with geometric shapes.''
Sam finds Picasso explored more with style in his late career, as several new clusters emerge toward the end of the dataset, including whimsical collages of musical instruments, minimalist minotaur sketches, and colorful still lifes.

\subsection{\textit{The Grand Budapest Hotel} Frames}
\begin{figure}
    \centering
    \includegraphics[width=\linewidth]{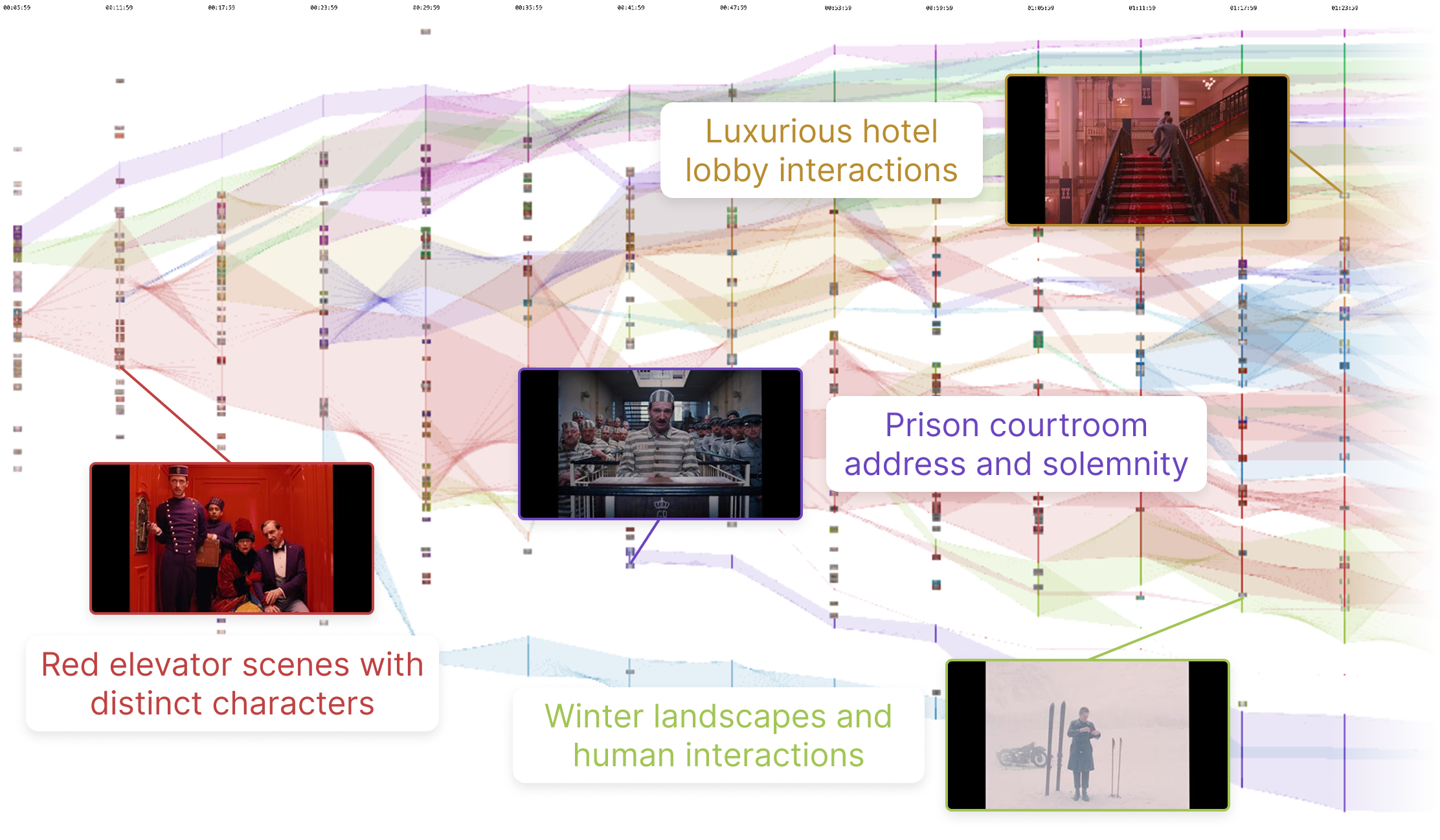}
    \caption{Topic clusters derived from \textit{The Grand Budapest Hotel} film frames visualized in 6-minute intervals using Side View.}
    \label{fig:budapest}
\end{figure}
Film aficionado Louis recently watched \textit{The Grand Budapest Hotel} and is curious about using \system to unpack visual themes across the film.
With \textit{Playback} and \textit{Front View}, Louis watches his favorite scenes unfold as he adds new data frames to the tool, such as when Gustave and Zero first meet at the hotel and the prison scene, which gets its own cluster (\coloredsquare{BPPurple} ``Prison courtroom address and solemnity'').
He finds that many clusters are defined by specific settings like ``hotel-elegant-lobby,'' but others are more general and characterized by subject and color like ``man-suit-purple,'' which makes sense given the film's rich cinematography.

Switching to \textit{Across} and labeling clusters with an LLM, Louis sees that the largest topic throughout the movie is ``Diverse social interactions in atmospheric settings,'' which branches into 11 child clusters including ``Dimly lit settings emphasizing human connection'' and \coloredsquare{BPRed} ``Red elevator scenes with distinct characters'' (Fig.~\ref{fig:budapest}). 
Another recurrent theme is ``Elegant hotel ambiance and interactions,'' which includes child topics such as \coloredsquare{BPYellow} ``Luxurious hotel lobby interactions'' and ``Urgent activity in opulent hotel settings.''
Later in the film, there are emergent topics such as \coloredsquare{BPGreen} ``Winter landscapes and human interactions,'' each representing a distinct aesthetic compared to existing clusters.

\section{Conclusion}
We introduce a technique that integrates force-based dimensionality reduction with streaming clustering to produce an interpretable visualization of how data changes over time. 
With it, we create \system, an interactive tool for exploring evolving topic trajectories and semantic patterns in time-based datasets.
Through use cases on text and images, we demonstrate how \system enables users to surface rich, temporally grounded insights that would be difficult to observe in static embeddings -- illustrating the co-evolution of semantic meaning and visual structure.
We hope our work sparks new ideas for dynamic temporal embedding interfaces.

\acknowledgments{
We are grateful for the feedback and support provided by our classmates in SCI 6492 and the members of the Harvard Insight + Interaction Lab in shaping this work.
We also thank the anonymous reviewers for their helpful comments and suggestions.

This work was supported by CY's National Science Foundation Graduate Research Fellowship under Grant No. DGE 2140743 and Kempner Institute Graduate Research Fellowship.
MW and FV received support from Effective Ventures Foundation, Effektiv Spenden Schweiz, an Open AI Superalignment grant, and the Open Philanthropy Project.
}

\bibliographystyle{abbrv-doi}

\bibliography{references}
\end{document}